# Cognitive styles sex the brain, compete neurally, and quantify deficits in autism

Nigel Goldenfeld[1], Sally Wheelwright[2] and Simon Baron-Cohen[2]

[1]*Department of Applied Mathematics and Theoretical Physics, Centre for Mathematical Sciences, University of Cambridge, Wilberforce Road, Cambridge CB3 0WA, England*

*and*

*Department of Physics, University of Illinois at Urbana-Champaign, 1110 West Green Street, Urbana, IL 61801, USA.*

[2]*Autism Research Centre, Departments of Experimental Psychology and Psychiatry, University of Cambridge, Douglas House, 18b Trumpington Road, Cambridge CB2 2AH, England.*

**Abstract**

*Introduction:* Two key dimensions of the mind are understanding and responding to another's mental state (empathizing), and analysing lawful behaviour (systemizing).

*Methods:* Two questionnaires, the Systemizing Quotient (SQ) and the Empathy Quotient (EQ), were administered to a normal control group and a group of individuals with Asperger Syndrome (AS) or High-Functioning Autism (HFA). The multivariate correlations of the joint scores were analysed using principal components analysis.

*Results:* The principal components were well-approximated by the sums and differences of the SQ and EQ scores. The differences in the scores corresponded to sex differences within the control group and also separated out the AS/HFA group, which showed stronger systemizing than the control group, but below-average empathy. The sums of the scores did not show sex differences, but did distinguish the AS/HFA group.

*Conclusions:* These tests reliably sex the brain, and their correlations show that empathizing and systemizing are not independent, but compete neurally. Their combined score (EQ + SQ) quantifies the deficit in autism spectrum conditions.

Two key modes of thought are systemizing and empathizing[1]. Systemizing is the drive to understand the rules governing the behaviour of a system and the drive to construct systems that are lawful. Systemizing allows one to predict and control such systems. Empathizing is the drive to identify another person's thoughts or emotions, and to respond to their mental states with an appropriate emotion. Empathizing allows one to predict another person's behaviour at a level that is accurate enough to facilitate social interaction. A growing body of data suggests that, on average, females are better than males at empathizing, and males are better than females at systemizing[2,3]. Here, we show that these abilities strongly differentiate brain sex, and moreover compete neurally, so that despite sex differences in cognitive style, there is no overall sex difference in cognitive ability.

Individuals with autism spectrum conditions have severe social difficulties and an 'obsessional' pattern of thought and behaviour[4]. Such diagnostic features may arise as a result of their significant disabilities in empathizing[5-7] as well as their stronger drive to systemize[8,9]. Such a cognitive profile, together with significant sex bias in incidence rate, is compatible with the theory that autism is an extreme of the male brain[1,10]. Our findings show that this is consistent with the cognitive style of individuals with autism spectrum conditions, and that there is a significant deficit in cognitive ability associated with the imbalance of cognitive style.

In order to quantify systemizing and empathizing, two self-report questionnaires have been developed[11]: the Systemizing Quotient (SQ) and the Empathy Quotient (EQ). In that study, these two questionnaires were tested in two groups: Group 1 comprised 114 males and 163 females randomly selected from the general population. Group 2 comprised 33 males and 14 females diagnosed with Asperger Syndrome (AS) or high-functioning autism (HFA). The mean scores of this study confirmed both the sex-difference in the general population (i.e., a male superiority in systemizing and a female superiority in empathizing), and the extreme male brain theory of autism.

**Method and Results**
Full details about the construction of the SQ and EQ questionnaires are available elsewhere [11,14]. Both were designed to be short, easy to complete, and easy to score. They have a forced-choice format, and are self-administered. Both the SQ and EQ comprise 60 questions, 40 assessing systemizing or empathizing (respectively), and 20 filler (control) items. Approximately half the items are worded to produce a "disagree" response, and half an "agree" response, for the systemizing/empathizing response. This is to avoid a response bias either way. Items are randomised. An individual scores 2 points if they strongly display a systemizing/empathiszing response, and 1 point if they slightly display a systemizing/empathizing response.

In the present study, we re-analysed the data reported in the earlier study[11] to test for a correlation between the scores for each individual on these tests. The maximum score on both questionnaires was 80. We plotted the raw scores from all individuals (from both groups) on a single chart, whose axes were labelled by the SQ and EQ scores, as shown in Figure 1a.

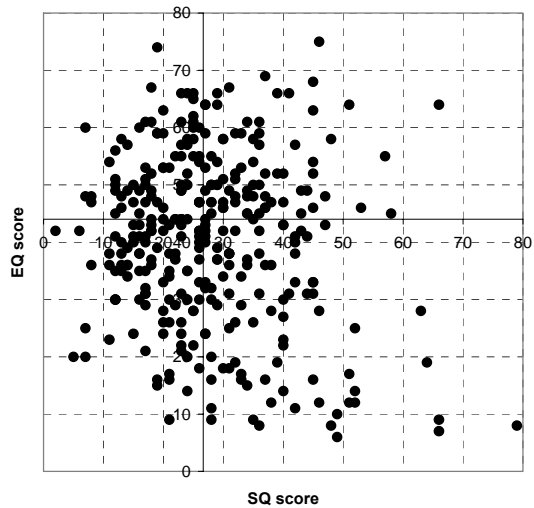 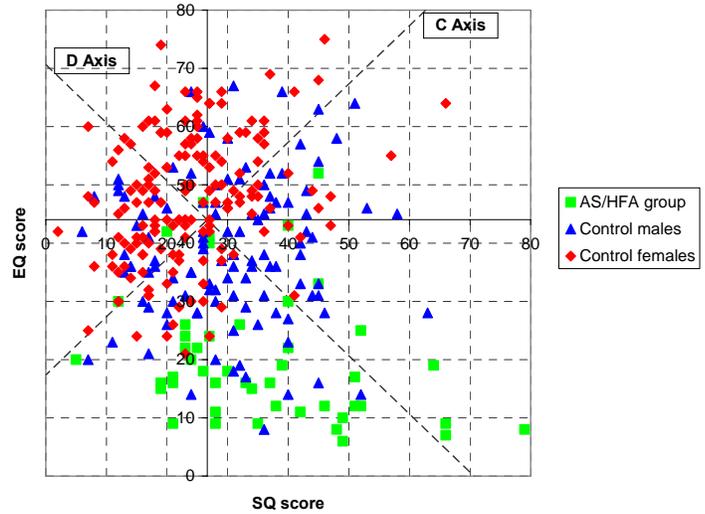

**Fig 1a:** SQ scores versus EQ scores for all participants. Note that the origin of the graph is at the controls' mean SQ and EQ scores.

**Fig 1b:** SQ scores versus EQ scores for all participants, separated into the 3 groups. Note that the origin of the graph is at the controls' mean SQ and EQ scores. Also shown are the C axis (the combined EQ and SQ scores) and the D axis (the difference between the SQ and EQ scores).

The means of each test were taken from Group 1 in the earlier data set, and in this way represent a sex-blind mean of the general population. As can be seen, the results cluster in the SQ-EQ space and do not randomly fill the chart. This suggests that it may not be possible to score anywhere in SQ-EQ space, and that there may be constraints operating, such that SQ and EQ are not independent.

We separated out the scores from the three groups: males from the general population (henceforth, male controls), females from the general population (female controls), and individuals with AS/HFA, as shown in colour in Figure 1b. Inspection of this plot strongly suggests 3 distinct populations. To explore the variations around the mean, we transformed the raw SQ and EQ scores into the two new variables: $S \equiv (SQ - <SQ>) / 80$ and $E \equiv (EQ - <EQ>) / 80$, i.e. we first subtracted the control population mean (denoted by $<...>$) from the scores, then divided by the maximum possible score, 80. The means were: 26.66 (SQ) and 44.01 (EQ). To reveal the differences between the populations we essentially factor analysed the results by performing a rotation of the original SQ and EQ axes by 45°. We normalised by the factors of ½ as is appropriate for an axis rotation. These new variables are defined as follows:

D = (S - E) / 2 (i.e., the difference between the normalised SQ and EQ scores) and
C = (S + E) / 2 (i.e., the sum of the normalised SQ and EQ scores).

D scores represent the difference in ability at systemizing and empathizing for each individual. A high D score can be attained either by being good at systemizing or poor at empathizing, or both. C scores test if systemizing and empathizing stand in a reciprocal, competitive relationship with each other, such that as one scores higher on one of these dimensions, one scores lower on the other. Competition might arise at the neural level (since space is limited in the cortex[12]) or might arise because both depend on some other biological resource (e.g., the hormone foetal testosterone[13]). If systemizing and empathizing are reciprocal, one would expect no difference in C scores between the sexes. These new D and C axes are shown in dotted lines on Figure 1b.

Figure 1b shows that the data have approximate boundaries that lie parallel to the C axis; in other words, the data vary significantly along the D dimension, but much less so along the C dimension. Our rotation was chosen to exhibit precisely this feature, but what was unexpected was that the rotation of 45° had such a natural interpretation, as explained below. Figure 1b suggests that the male control data have greater weight than the female data on the positive D axis, and the AS/HFA group has weight even further to the right along that axis than the male controls. By contrast, there is no significant trend along the C axis.

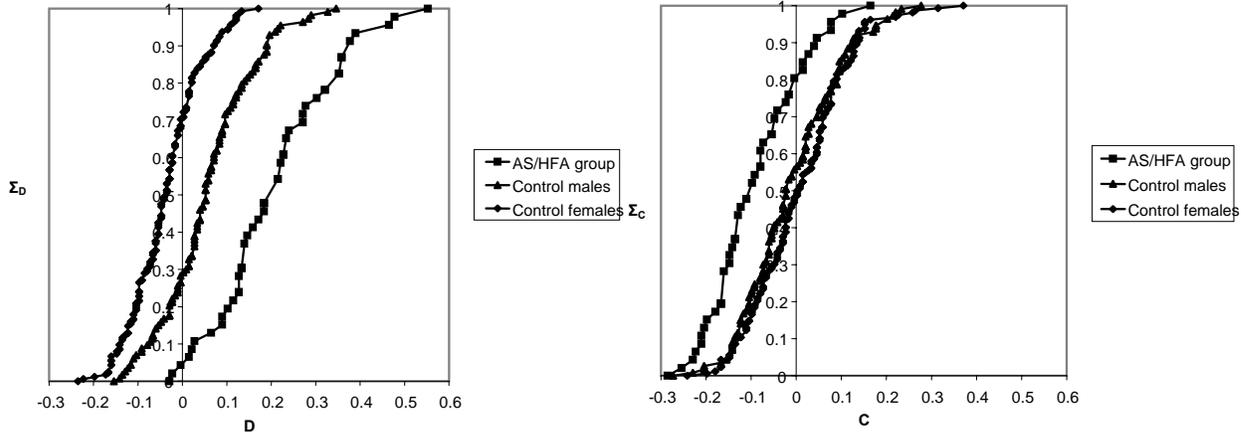

**Fig 2a: Cumulative distribution function ($\Sigma_D$) of D**  **Fig 2b: Cumulative distribution function ($\Sigma_C$) of C**

To explore this further, we plotted the cumulative distributions of our data along the D and C directions, making separate plots for control male, control female and AS/HFA groups. We define the cumulative distribution $\Sigma_D(D)$ along the D direction as the fraction of data points whose D value is less than D' irrespective of the C value (see Figure 2a).

Similarly, we define the cumulative distribution $\Sigma_C(C)$ along the C direction as the fraction of data points whose C value is less than C', irrespective of the D value (see Figure 2b).

The means and standard deviations of the C and D scores for the different populations are as follows: D scores: control females = -0.039 (0.006); control males = 0.055 (0.011); AS/HFA = 0.21 (0.018). C scores: control females = 0.007 (0.011); control males = -0.0 (0.012); AS/HFA = -0.092 (0.010).

Figure 2a shows the cumulative distribution along the D direction, $\Sigma_D$, plotted for the three different groups: control female, control male and AS/HFA. The cumulative distributions are widely spaced apart, much further than the fluctuations in the raw data, indicating that these groups really do represent three distinct populations and are not sampled from the same underlying distribution. We quantified this observation by performing a between-subjects single-factor analysis of variance (ANOVA). There was a significant effect of group ($F(2, 321) = 121$, $p < 0.0001$). Post-hoc Tukey tests confirmed that all 3 groups differed significantly from one another.

Figure 2b shows the cumulative distribution along the C direction, $\Sigma_C$, plotted for the three different groups: control female, control male and AS/HFA. It is apparent that the control male and control female plots are indistinguishable up to the sample fluctuations, but both are well separated from the plots for the AS/HFA group. We quantified this observation by performing a between-subjects single-factor analysis of variance (ANOVA). As expected, there was a significant effect of group ($F(2, 321) = 16.2$, $p < 0.0001$). Post-hoc Tukey tests confirmed that there was no significant difference between control males and females, but both of these groups were significantly different from the AS/HFA group.

These results indicate that the control male and female groups show distinct and significant differences in their cognitive style. The male group scores higher than the female group along the D dimension (relatively higher systemizing and lower empathizing), but there is no difference between the sexes in the measure of C (combined scores). Apparently, females' relatively high empathizing ability compensates for their less developed systemizing ability, and conversely males' high systemizing ability compensates for their less well-developed empathizing skills. The AS/HFA group has a lower C score. This is because, although they outperform both male and female controls on the systemizing measure, this does not compensate for their much lower scores on the empathizing measure.

Previously, a classification of brain types was proposed[1], based in part on the empirical evidence suggesting that, as a group, males score higher on the SQ, but lower on the EQ, relative to females[11]. These data also suggested the possibility of a weak inverse relation between SQ and EQ scores. This inverse relationship is fully exposed by the analysis presented here. In particular, because the sex-differences are only discernable along the D dimension, regions of similar brain type are bounded by lines that are parallel to the C axis, or in terms of the original raw data, lines that lie parallel to the lower-left to upper-right diagonal of the SQ-EQ plot. Since there is no unique way to break up the results of our data analysis into identifiable groups along the D dimension, we propose a classification based upon the cumulant plot of Figure 2a. This generates 5 brain types, as follows:

(1) A significant proportion of individuals in the general population is likely to have a 'balanced' brain (or be of Type B), that is, their E and the S are not significantly different to each other. This can be expressed as E ≈ S. In practice, we defined this as individuals whose D score lay between the median of the control male and female populations.

(2) A proportion of the general population is likely to have an 'extreme S' Type brain, that is, having a D score larger than the median of the AS/HFA group. This can be expressed as S>>E.

(3) A proportion of the general population is likely to have an 'extreme E' Type brain, symmetrically opposite to the extreme S Type brain. This can be expressed as E>>S. (We are not aware of any known clinical group which corresponds to this).

(4) The S Type brain can then be defined as those individuals who lie between the Type B and the extreme Type S brains. This can be expressed as S>E.

(5) The E Type brain can then be defined as those individuals who lie between the Type B and the extreme Type E brains. This can be expressed as E>S.

These 5 brain type definitions are based upon median scores, rather than *a priori* criteria based upon the mean and standard deviation. This obviates the need to make special assumptions about the form of the distributions. Table 1 shows the percentage of each of the 3 groups of individuals falling into each of the 5 Types of brain, using the median definitions above.

Table 1 also shows that similar results were obtained by using a classification based upon the control males and females and simply taking a range of percentiles that separated out the tails of the distribution and the center.

These natural groupings can be defined in terms of the deviations of the SQ and EQ scores from the means over the control populations. Thus, the balanced (B) brain type refers to individuals whose scores are close to the respective means, while S and E are brain types where the deviation from the mean is much greater in S (E) than for E (S). Similarly, extreme S and extreme E are extreme forms of brain types S and E respectively.

| Brain Type | Extreme E | E | B | S | Extreme S |
|---|---|---|---|---|---|
| Brain Sex | Extreme female | Female | Balanced | Male | Extreme male |
| Defining Characteristic | S << E | S < E | S ≈ E | S > E | S >> E |
| Brain types based on median positions of the three sub-populations male, females, AS/HFA | | | | | |
| Brain Boundary (median) | D < -0.16 | -0.16 < D < 0.035 | -0.035 < D < 0.052 | 0.052 < D < 0.21 | D > 0.21 |
| Female % | 7 | 47 | 32 | 14 | 0 |
| Male % | 0 | 17 | 31 | 46 | 6 |
| AS/HFA % | 0 | 0 | 13 | 40 | 47 |
| Brain types based on percentiles of male and female controls | | | | | |
| Brain Boundary (percentile) | D < -0.16 | -0.16 < D < -0.048 | -0.048 < D < 0.027 | 0.027 < D < 0.21 | D > 0.21 |
| percentile (per) | per < 2.5 | 2.5 ≤ per < 35 | 35 ≤ per < 65 | 65 ≤ per < 97.5 | per ≥ 97.5 |
| Female % | 4.3 | 44.2 | 35.0 | 16.5 | 0 |
| Male % | 0 | 16.7 | 23.7 | 53.5 | 6.1 |
| AS/HFA % | 0 | 0 | 12.8 | 40.4 | 46.8 |

**Table 1. Classifications of brain type based upon median positions of the sub-populations control males, females and AS/HFA (data from figure 2a), and upon percentiles of the entire sample (data from figure 1a). Both classifications give similar results.**

With the median definitions as given in Table 1, we note that there are significant sex differences in the populations of the different brain types. In the balanced brain type, males and females are present in virtually equal proportions. However, in S-type brains, males outnumber females by a factor of nearly 3:1. In E-type brains, females outnumber males by about the same factor. Finally, among the extreme S-type brains, individuals diagnosed with AS/HFA outnumber males by a factor of nearly 10. Unfortunately, there are not enough data to make any determination of sex-related trends within the AS/HFA group. We hope that future studies will be able to address this interesting question. These trends, rather than the precise boundaries we have chosen between the brain types, are the key differences that our SQ and EQ studies expose, and are not very sensitive to whether the median or percentile classification is used.

In order to present these results in a practical form, we show in Figure 3 our results for the different brain types (using the median definitions), translated back into raw scores on the SQ and EQ tests. Figure 3 can be directly used to classify an individual's brain type as represented by their responses to the SQ and EQ tests.

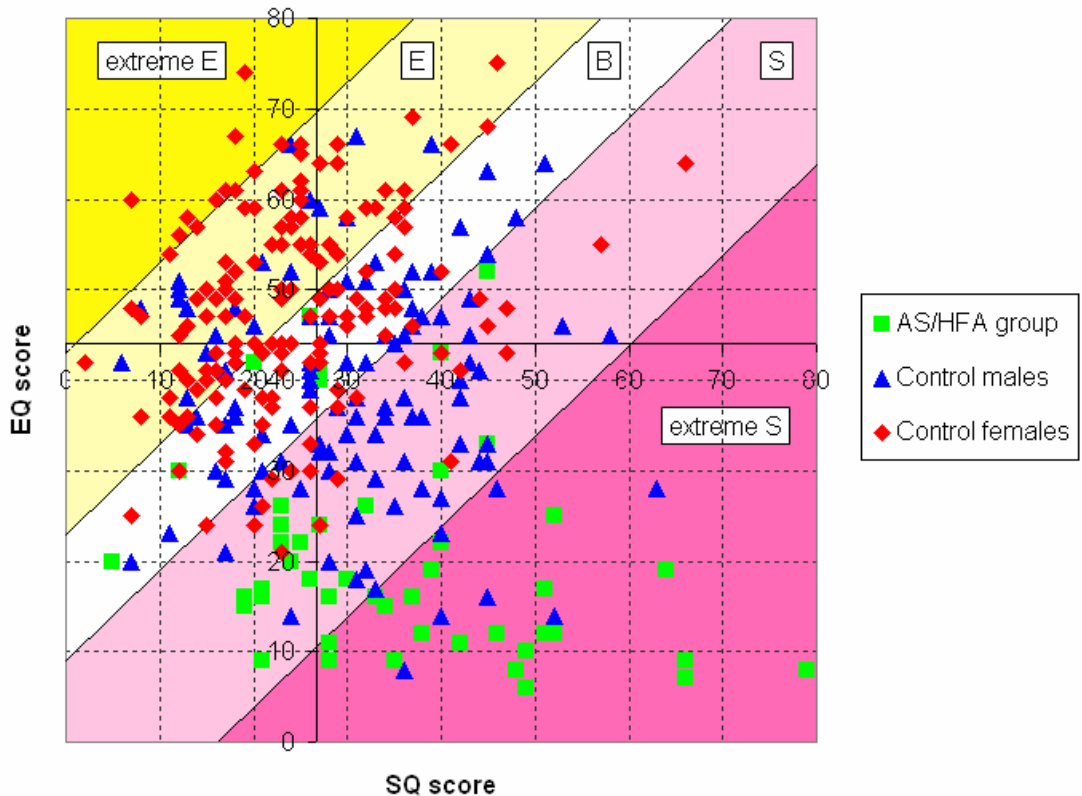

**Fig 3: SQ scores versus EQ scores for all participants with the proposed boundaries for the different brain types**

**Conclusions**

We have shown that a re-analysis of the data from an earlier study using the Empathy Quotient (EQ) and Systemizing Quotient (SQ)[11] reliably sexes the brain when analysed blind. In addition, although females show stronger empathizing and males show stronger systemizing, their *combined* scores do not differ, suggesting that empathizing and systemizing compete neurally in the brain. This also suggests that overall, neither sex is superior. We also confirm earlier reports that people with Asperger Syndrome (AS) or high functioning autism (HFA) have stronger systemizing scores than normal, but our new analysis shows that this did not compensate for their weaker E: thus their combined scores do not equal those of the normal groups. This result lends support to the extreme male brain theory of autism, and confirms that autism spectrum conditions arise from a cognitive deficit.

**Acknowledgments**: We thank F. Durbridge for helpful comments. SBC and SW were supported by the Medical Research Council of the United Kingdom. NG was supported by the US National Science Foundation during the period of this work.

**Corresponding author**: Please address communication regarding this article to Nigel Goldenfeld at nigel@uiuc.edu.

**Human subjects:** Informed consent was obtained from the subjects after the nature and possible consequences of the study were explained.